\documentstyle[12pt,tighten,aps,psfig]{revtex} 
\def\be{\begin{equation}}
\def\e#1{\label{#1}\end{equation}}
\def\r#1{(\ref{#1})}

\begin{document}
\draft
\title{The Zeno and anti-Zeno effects on decay in dissipative quantum 
systems\footnote{
Published in acta physica slovaca {\bf 49}, 541-548 (1999),
under the title ``Decay control in dissipative quantum systems''
}}
\author{A. G. Kofman\footnote{E-mail address:
abraham.kofman@weizmann.ac.il} and G. Kurizki\footnote{E-mail address:
gershon.kurizki@weizmann.ac.il}}
\address{Department of Chemical Physics, The Weizmann Institute of 
Science, Rehovot 76100, Israel}
\date{Received 10 May 1999, accepted 12 May 1999}
\maketitle
\begin{abstract}
We point out that the quantum Zeno effect, i.e., inhibition of 
spontaneous decay by frequent measurements, is observable only in 
spectrally finite reservoirs, i.e., in cavities and waveguides, using 
a sequence of evolution-interrupting pulses or randomly-modulated CW
fields. By contrast, such measurements can only accelerate decay in
free space. 
\end{abstract}

\section{Introduction}

The "watchdog" or quantum Zeno effect (QZE) is a basic
manifestation of the influence of measurements on the evolution of a
quantum system. The original QZE prediction has been that {\em 
irreversible decay} of an excited state into an open-space reservoir 
can be inhibited \cite{4},
by repeated interruption of the system-reservoir coupling, which is
associated with measurements (e.g., the interaction of an unstable
particle with its environment on its flight through a bubble chamber) 
\cite{5,6}. However, this prediction has
not been experimentally verified as yet! Instead, the interruption of
Rabi oscillations and analogous forms of {\em nearly-reversible\/}
evolution has been at the focus of interest 
\cite{7,8,8a,8b,8c,8d,9,10}. Tacit assumptions have been made that
the QZE is in principle attainable in open space, but is technically
difficult.

We have recently demonstrated \cite{11} that the inhibition of {\em
nearly-exponential} excited-state decay by the QZE in two-level atoms,
in the spirit of the original suggestion \cite{4}, is amenable to
experimental verification in resonators. Although this task has been
widely believed to be very difficult, we have shown, by means of our
unified theory of spontaneous emission into arbitrary reservoirs 
\cite{12}, that two-level emitters in cavities or in waveguides are 
in fact adequate for radiative decay control by the QZE\cite{11}. 
Condensed media or multi-ion traps are their analogs for vibrational 
decay control (phonon emission) by the QZE\cite{har98}. We have now 
developed a more comprehensive view of the possibilities of 
excited-state decay by QZE. Here we wish to demonstrate that QZE is 
indeed achievable by repeated or continuous measurements of the 
excited state, but only in reservoirs whose spectral response {\em 
rises up} to a frequency which does not exceed the resonance
(transition) frequency. By contrast, in 
open-space decay, where the reservoir response has a much higher
cutoff, {\em non-destructive} frequent measurements are much more
likely to accelerate decay, causing the anti-Zeno effect.

\section{Measurement schemes}

\subsection{Impulsive measurements (Cook's scheme)}

Consider an initially excited two-level atom coupled to an {\em
arbitrary} density-of-modes (DOM) spectrum $\rho(\omega)$ of the
electromagnetic field in the vacuum state. At time $\tau$ its
evolution is interrupted by a short optical pulse, which serves as an
impulsive quantum measurement \cite{7,8,8a,8b,8c,8d,9,10}. Its role is
{\em to break the evolution coherence}, by transferring the
populations of the excited state $|e\rangle$ to an auxiliary state
$|u\rangle$ which then decays back to $|e\rangle$ {\em incoherently}.

The spectral response, i.e., the emission rate into this reservoir at
frequency $\omega$, is 
\begin{equation}
G(\omega)=|g(\omega)|^2\rho(\omega),
\label{1}
\end{equation}
$\hbar g(\omega)$ being the field-atom coupling energy.

We cast the excited-state amplitude in the form
$\alpha_e(\tau)e^{-i\omega_a\tau}$, where $\omega_a$ is the atomic
resonance frequency.  Restricting ourselves to sufficiently short
interruption intervals $\tau$ such that $\alpha_e(\tau)\simeq 1$, yet
long enough to allow the rotating wave approximation, we obtain
\begin{eqnarray}
\alpha_e(\tau)&\simeq&1-\int_0^\tau dt(\tau-t)\Phi(t)e^{i\Delta t},
\label{3}
\end{eqnarray}
where
\begin{equation}
\Phi(t)=\int_0^\infty d\omega G(\omega)e^{-i(\omega-\omega_s)t}.
\label{2b}
\end{equation}
$\Delta=\omega_a-\omega_s$ is the detuning of the atomic resonance
from the peak (or cutoff) $\omega_s$ of $G(\omega)$.

To first order in the atom-field interaction, the excited state
probability after $n$ interruptions (measurements),
$W(t=n\tau)=|\alpha_e(\tau)|^{2n}$, can be written as
\begin{equation} 
W(t=n\tau)\approx[2\mbox{Re}\alpha_e(\tau)-1]^n\approx e^{-\kappa t},
\label{4'}\end{equation}
where
\begin{equation}
\kappa=\frac{2}{\tau}\mbox{Re}[1-\alpha_e(\tau)]
=\frac{2}{\tau}\mbox{Re}\int_0^\tau dt(\tau-t)\Phi(t)e^{i\Delta t}.
\label{4}
\end{equation}
The QZE obtains if $\kappa$
{\em decreases with} $\tau$ for sufficiently short $\tau$. This
essentially means that the correlation (or memory) time of the field
reservoir is longer (or, equivalently, $\Phi(t)$ falls off slower)
than the chosen interruption interval $\tau$.

Equation \r{4} can be rewritten as 
\be
\kappa=2\pi\int G(\omega)\left\{\frac{\tau}{2\pi}
\text{sinc}^2\left[\frac{(\omega-\omega_a)\tau}{2}\right]\right\}
d\omega,
\e{5}
where the interruptions are seen to cause dephasing whose spectral
width is $\sim 1/\tau$.

\begin{figure}
\begin{center}
\begin{tabular}{cc}
\psfig{file=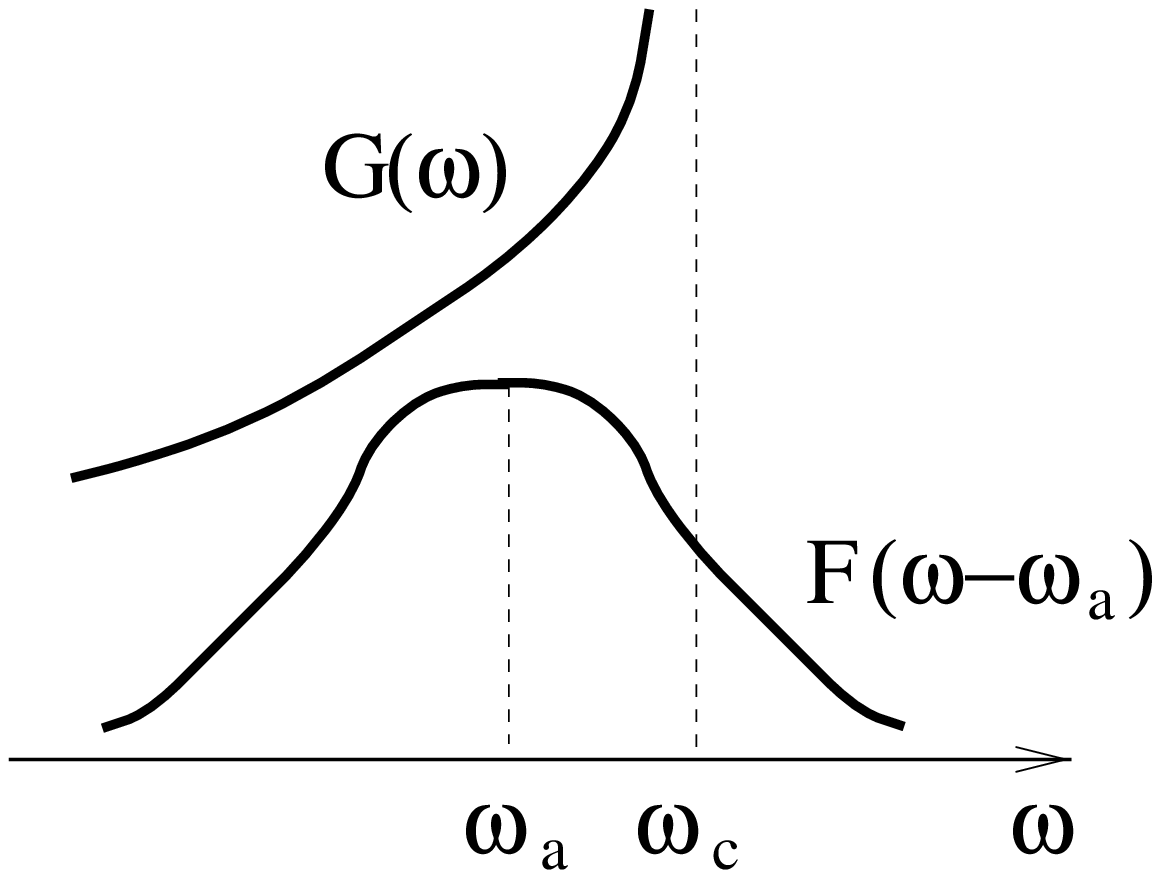,width=6cm}&
\ \ \ \ \psfig{file=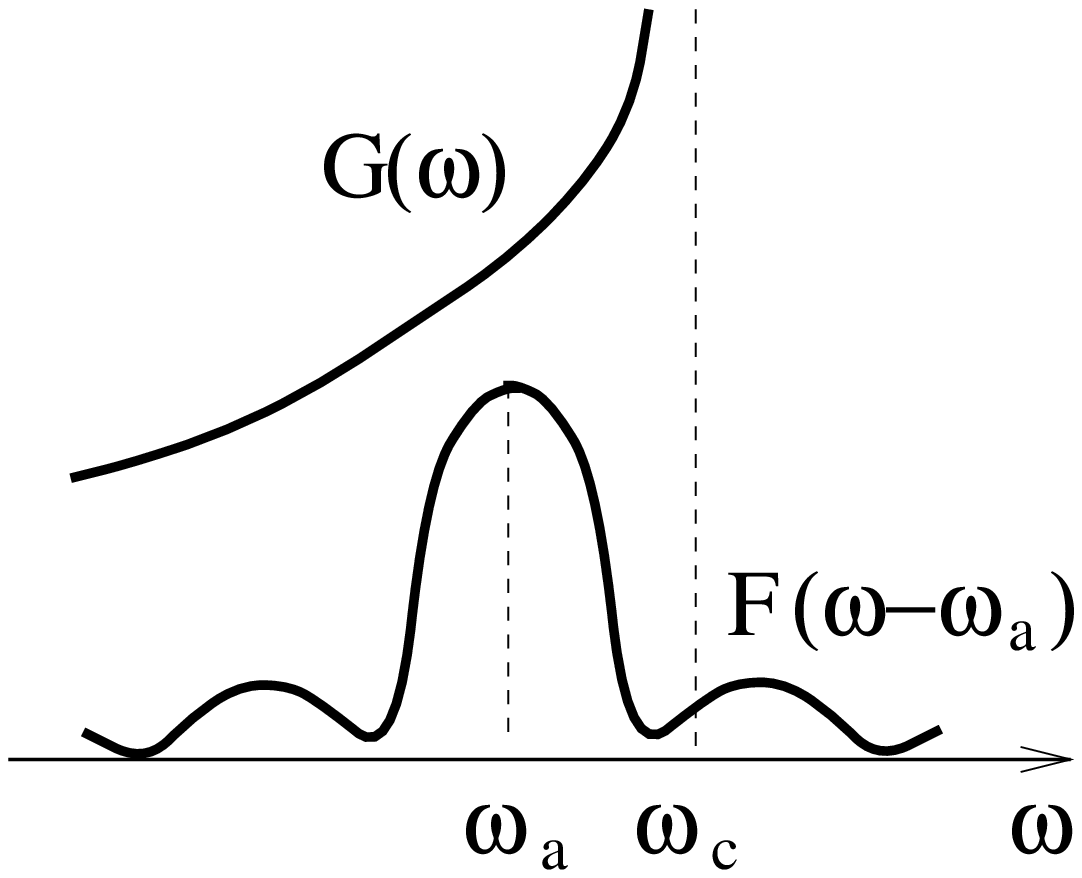,width=6cm}\\
(a) & (b)
\end{tabular}
\end{center}
\caption{  Dependence of effective decay rate $\kappa$
on dephasing (relaxation) spectrum
$F(\Delta)$ and field reservoir response with cutoff $G(\omega)$: (a)
Lorentzian dephasing spectrum [Eq. (\protect\ref{14})]; 
(b) sinc-function 
spectrum [Eq. (\protect\ref{5}) -- impulsive measurements].  }
\label{fig:4}\end{figure}

\subsection{Noisy-field dephasing: Random Stark shifts}

Instead of disrupting the coherence of the evolution by a sequence of
"impulsive" measurements, as above, we can achieve
this goal by {\em noisy-field dephasing} of $\alpha_e(t)$: Random
ac-Stark shifts by an off-resonant intensity-fluctuating field result
in the replacement of Eq. (\ref{5}) by (Fig. \ref{fig:4})
\begin{equation}
\kappa=\int G(\Delta+\omega_a){\cal L}(\Delta)d\Delta,
\label{14}
\end{equation}
Here the spectral response $G(\Delta+\omega_a)$ is the same as in
Eq. (\ref{1}), whereas ${\cal L}(\Delta)$ is the Lorentzian-shaped
relaxation function of the coherence element $\rho_{eg}(t)$, which 
for the common dephasing model decays {\em exponentially}. This
Lorentzian relaxation spectrum has a HWHM width 
$\nu=\langle\Delta\omega^2\rangle\tau_c$, the product of the 
mean-square
Stark shift and the noisy-field correlation time. The QZE condition is
that {\em this width be larger than the width of\/} $G_s(\omega)$
(Fig. \ref{fig:4}). The advantage of this realization is that it does
not depend on $\gamma_u$, and is realizable for any atomic
transition. Its importance for molecules is even greater: if we start
with a single vibrational level of $|e\rangle$, no additional levels
will be populated by this process.

\subsection{CW dephasing}

The random ac-Stark shifts described above cause both shifting and
broadening of the spectral transition. If we wish to avoid the
shifting altogether, we may employ a CW
driving field that is {\em nearly resonant} with the
$|e\rangle\leftrightarrow|u\rangle$ transition\cite{7,8}. If the 
decay rate of this transition, $\gamma_u$, is larger than the Rabi
frequency $\Omega$ of the driving field, then one can show that
$\kappa$ is given again by Eq. \r{14}, where the Lorentzian
(dephasing) width is
\begin{equation}
\nu=\frac{2\Omega^2}{\gamma_u}. 
\label{15}\end{equation}

\subsection{Universal formula}

All of the above schemes are seen to yield the same universal
formula for the decay rate
\be
\kappa=2\pi\int G(\omega)F(\omega-\omega_a)d\omega,
\e{8}
where $F(\omega)$ expresses the relevant measurement-induced 
dephasing (sinc- or a Lorentzian-shaped): its
width relative to that of $G(\omega)$ determines the QZE behavior.

\begin{figure}
{\vspace*{-3cm}
\centerline{\psfig{file=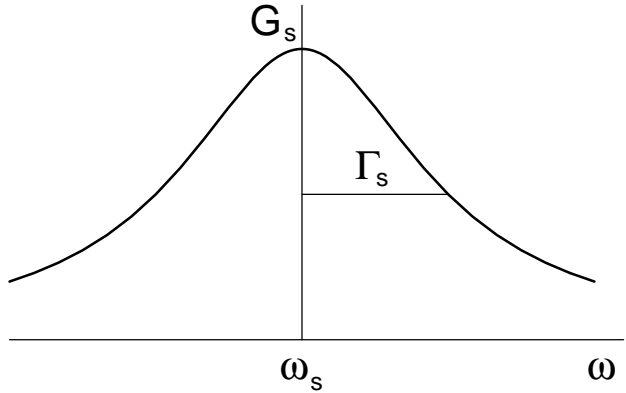,width=3.375in}}
\vspace{-3cm}}
\caption{ Cavity mode with Lorentzian lineshape.
}\label{fig:1}\end{figure}

\section{Applications to various reservoirs}

\subsection{Finite reservoirs: A Lorentzian line}

The simplest application of the above analysis is to the case of a
two-level atom coupled to a near-resonant Lorentzian line centered at
$\omega_s$, characterizing a high-$Q$ cavity mode \cite{11}. In this
case, 
\be
G_s(\omega)=\frac{g_s^2\Gamma_s}{\pi[\Gamma_s^2+(\omega-\omega_s)^2]},
\e{9}
where $g_s$ is the resonant coupling
strength and $\Gamma_s$ is the linewidth (Fig. \ref{fig:1}). 
Here $G_s(\omega)$ stands for the sharply-varying (nearly-singular)
part of the DOM distribution, associated with narrow cavity-mode lines
or with the frequency cutoff in waveguides or photonic band edges. The
broad portion of the DOM distribution $G_b(\omega)$ (the "background" 
modes), always coincides with
the free-space DOM $\rho(\omega)\sim\omega^2$ at frequencies well
above the sharp spectral features. In an open cavity,
$G_b(\omega)$ represents the atom coupling to the unconfined
free-space modes. This
gives rise to an {\em exponential} decay factor in
the excited state probability, regardless of how short $\tau$ is,
i.e.,
\begin{equation}
\kappa=\kappa_s+\gamma_b,
\label{4a}
\end{equation}
where $\kappa_s$ is the contribution to $\kappa$ from the
sharply-varying modes and
$\gamma_b=2\pi G_b(\omega_a)$ is the effective rate of 
spontaneous emission into the background modes.
In most structures $\gamma_b$ is comparable to the free-space decay
rate $\gamma_f$.

In the
short-time approximation, taking into account that the Fourier
transform of the Lorentzian $G_s(\omega)$ is $\Phi_s(t) =
g_s^2e^{-\Gamma_s t}$, Eq. (\ref{3}) yields (without the
background-modes contribution)
\begin{equation}
\alpha_e(\tau)\approx 1-\frac{g_s^2}{\Gamma_s-i\Delta}\left[
\tau+\frac{e^{(i\Delta-\Gamma_s)\tau}-1}{\Gamma_s-i\Delta}\right].
\label{6}
\end{equation}
The QZE condition is then
\begin{equation}
\tau\ll(\Gamma_s+|\Delta|)^{-1},g_s^{-1}.
\label{7a}\end{equation}
{\em On resonance}, when $\Delta=0$, Eqs. \r{4} and (\ref{6}) yield
\begin{equation}
\kappa_s=g_s^2\tau.
\label{7}
\end{equation}

Thus the background-DOM effect cannot be modified by QZE. Only the
sharply-varying DOM contribution $\kappa_s$ may allow for QZE.
Only the $\kappa_s$ term decreases with $\tau$, indicating the QZE
inhibition of the nearly-exponential decay into the Lorentzian field
reservoir as $\tau\rightarrow 0$.  Since $\Gamma_s$ has dropped out of
Eq.~(\ref{7}), the decay rate $\kappa$ is the {\em same} for both
strong-coupling ($g_s > \Gamma_s$) and weak-coupling ($g_s \ll
\Gamma_s$) regimes. Physically, this comes about since for $\tau\ll
g_s^{-1}$ the energy uncertainty of the emitted photon is too large to
distinguish between reversible and irreversible evolutions.

The evolution inhibition, however, has rather different meaning for
the two regimes. In the weak-coupling regime, where, in the absence of
the external control, the excited-state population decays nearly
exponentially at the rate $g_s^2/\Gamma_s+\gamma_b$ (at $\Delta=0$),
one can speak about the inhibition of irreversible decay, in the
spirit of the original QZE prediction \cite{4}. By contrast, in the
strong-coupling regime in the absence of interruptions (measurements),
the excited-state population undergoes damped Rabi oscillations at the
frequency $2g_s$. In this case, the QZE slows down the evolution
during the first Rabi half-cycle ($0 \leq t \leq \pi / 2 g_s^{-1}$),
the evolution on the whole becoming irreversible.

\begin{figure}
{\centerline{\psfig{file=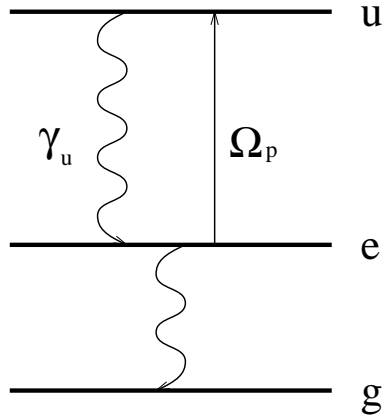,width=2.in}}}
\caption{Cook's scheme for impulsive measurements.
\label{f5} }
\end{figure}

A possible realization of this scheme is as follows. Within an open
cavity the atoms repeatedly interact with a pump laser, which is
resonant with the $|e\rangle\rightarrow|u\rangle$ transition
frequency. The resulting $|e\rangle\rightarrow|g\rangle$ fluorescence
rate is collected and monitored as a function of the pulse repetition
rate $1/\tau$. Each short, intense pump pulse of duration $t_p$ and
Rabi frequency $\Omega_p$ is followed by spontaneous decay from
$|u\rangle$ back to $|e\rangle$, at a rate $\gamma_u$, so as to {\em
destroy the coherence} of the system evolution, on the one hand, and
{\em reshuffle the entire population} from $|e\rangle$ to $|u\rangle$
and back, on the other hand (Fig. \ref{f5}). The demand that the
interval between measurements significantly exceed the measurement
time, yields the inequality $\tau\gg t_p$. The above inequality can be
reduced to the requirement $\tau\gg\gamma_u^{-1}$ if the
``measurements'' are performed with $\pi$ pulses: $\Omega_pt_p=\pi,\
t_p\ll\gamma_u^{-1}$. This calls for choosing a
$|u\rangle\rightarrow|e\rangle$ transition with a much shorter
radiative lifetime than that of $|e\rangle\rightarrow|g\rangle$.

\begin{figure}
{\vspace*{-3cm}
\centerline{\psfig{file=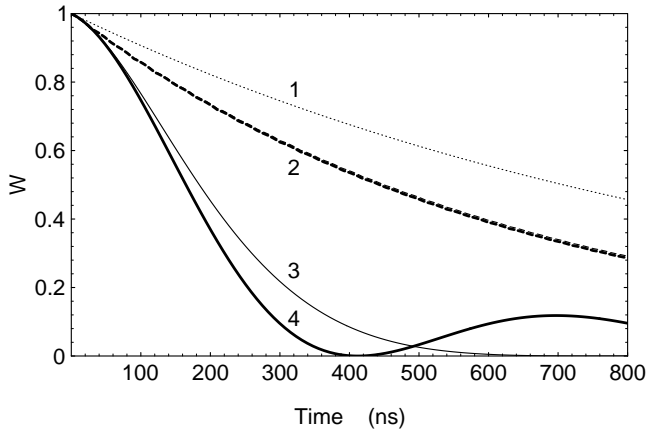,width=3.375in}}
\vspace{-3cm}}
\caption{
Evolution of excited-state population $W$ in
two-level atom coupled to cavity mode with Lorentzian lineshape on
resonance ($\Delta=0$): 
curve 1---decay to background-mode continuum
at rate $\gamma_b\simeq\gamma_f=10^6$ s$^{-1}$; curve
3---uninterrupted decay in cavity with $F\equiv(1-R)^{-2}=10^5$,
$L$=15 cm, and $f$=0.02 ($\Gamma_s=6.3\times 10^6$ s$^{-1}$, 
$g_s=4.5\times 10^6$ s$^{-1}$); curve 4---idem, but with 
$F=10^6$ ($\Gamma_s=2\times 10^6$ s$^{-1}$; damped 
Rabi oscillations); curve 2---interrupted evolution along {\em both}
curves 3 and 4, at intervals $\tau=3\times 10^{-8}$ s.
\label{f3} }
\end{figure}

\begin{figure}
{\vspace*{-3cm}
\centerline{\psfig{file=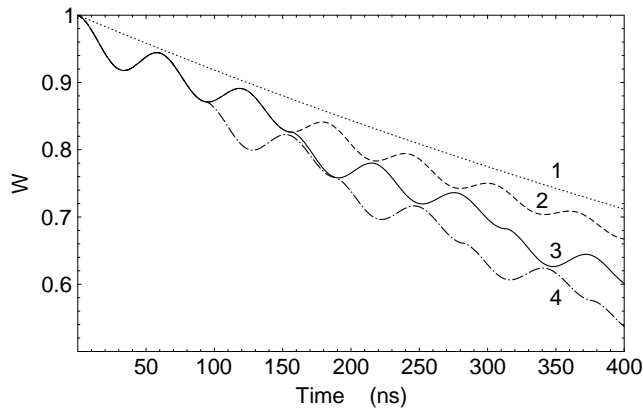,width=3.375in}}
\vspace{-3cm}}
\caption{Idem, for detuning $\Delta=10^8$s${}^{-1}$ and $F=10^6$:
curve 1---decay to background-mode continuum; 
curve 2---uninterrupted free evolution;
curve 3---interrupted evolution at intervals  
$\tau=5\pi\times 10^{-8}$ s ($\Delta\tau=5\pi$); curve 4---idem, for 
$\tau=3\pi\times 10^{-8}$ s ($\Delta\tau=3\pi$). 
\label{f4}}
\end{figure}

Figure \ref{f3}, describing the QZE for a Lorentz line on resonance
($\Delta=0$), has been programmed for feasible cavity parameters:
$\Gamma_s=(1-R)c/L,\ g_s=\sqrt{cf\gamma_f/(2L)},\ \gamma_b=(1-f)
\gamma_f$, where $R$ is the geometric-mean reflectivity of the two
mirrors, $f$ is the fractional solid angle (normalized to $4\pi$)
subtended by the confocal cavity, and $L$ is the cavity length. It
shows, that the population of $|e\rangle$ decays nearly-exponentially
well within interruption intervals $\tau$, but when those intervals
become too short, there is significant inhibition of the decay. 
Figure \ref{f4} shows the effect of the detuning 
$\Delta=\omega_a-\omega_s$ on the
decay: The decay now becomes oscillatory. The interruptions now {\em
enhance\/} the decay, the degree of enhancement depends on the phase
between interruptions.

\subsection{Open-space reservoirs}
The spectral response for hydrogenic-atom radiative decay via the 
$\vec{p}\cdot\vec{A}$ free-space interaction is given by\cite{mos72}
\be
G(\omega)=\frac{\alpha\omega}{[1+(\omega/\omega_c)^2]^4},
\e{10}
where $\alpha$ is the effective field-atom coupling constant and the 
cutoff frequency is
\be
\omega_{\rm c}\approx 10^{19}\ \text{s}^{-1}\sim\frac{c}{a_{\rm B}}.
\e{11}
Using measurement control that produces Lorentzian broadening [Eq.
\r{14}] we then obtain
\be
\kappa=\frac{\alpha\omega_{\rm c}}{3}\text{Re}\left[
\frac{f(2f^4-7f^2+11)}{2(f^2-1)^3}-\frac{6f\ln f}{(f^2-1)^4}
-\frac{3i\pi(f^2+4f+5)}{16(f+1)^4}\right],
\e{12}
where
\be
f=\frac{\nu-i\omega_a}{\omega_{\rm c}}.
\e{13}
In the range
\be
\nu\ll\omega_{\rm c}
\e{16}
we obtain from Eq. \r{12} the {\em anti-Zeno effect} of 
accelerated decay. This
comes about due to the {\em rising} of the spectral response
$G(\omega)\approx\alpha\omega$ as a function of frequency (for 
$\omega\ll\omega_{\rm c}$).
The Zeno effect can hypothetically occur only for 
$\nu\agt\omega_{\rm c}\sim 10^{19}$ s$^{-1}$.
But this range is well beyond the limit of validity of the present
analysis, since $\Delta E\sim\hbar\nu\agt\hbar\omega_{\rm c}$ may 
then induce other decay channels ("destruction") of $|e\rangle$, in
addition to spontaneous transitions to $|g\rangle$.

\section{Conclusions}

Our unified
analysis of two-level system coupling to field reservoirs has revealed
the general optimal conditions for observing the QZE in various
structures (cavities, waveguides, phonon reservoirs, and photonic band
structures) as opposed to open space. We note that the wavefunction 
collapse notion is not involved here, since the measurement is 
explicitly described as an act of dephasing (coherence-breaking). 
This analysis also clarifies that QZE cannot combat the open-space
decay. Rather, impulsive or continuous dephasing are much more likely
to accelerate decay by the inverse (anti-) Zeno effect.


\begin{references}

\bibitem{4} B. Misra and E. C. G. Sudarshan, ``The Zeno paradox in
quantum theory,'' J. Math. Phys. {\bf 18}, 756 (1977).

\bibitem{5} J. Maddox, ``Can observations prevent decay?'', 
\nat {\bf 306}, 111 (1983).

\bibitem{6} A.Peres, ``Quantum limited detectors for weak classical
signals,'' \prd {\bf 39}, 2943 (1989).

\bibitem{7} 
W. M. Itano, D. J. Heinzen, J. J. Bollinger, and D. J. Wineland, 
``Quantum Zeno effect,'' \pra {\bf 41}, 2295 (1990).

\bibitem{8} P. L. Knight, ``The quantum Zeno effect,'' \nat {\bf
344}, 493 (1990).

\bibitem{8a} T. Petrosky, S. Tasaki, and I. Prigogine, ``Quantum Zeno
effect,'' \pl A {\bf 151}, 109 (1990).

\bibitem{8b} E. Block and P. R. Berman, ``Quantum Zeno effect and
quantum Zeno paradox in atomic physics,'' \pra {\bf 44}, 1466 (1991).

\bibitem{8c} L. E. Ballentine, ``Quantum Zeno effect - comment,'' \pra
{\bf 43}, 5165 (1991).

\bibitem{8d} V. Frerichs and A. Schenzle, ``Quantum Zeno effect
without collapse of the wave packet,'' \pra {\bf 44}, 1962 (1991).

\bibitem{9} M. B. Plenio, P. L. Knight, and R. C. Thompson,
``Inhibition of spontaneous decay by continuous measurements -
proposal for realizable experiment,'' \oc {\bf 123}, 278 (1996).

\bibitem{10} A. Luis and J. Pe\v{r}ina, ``Zeno effect in parametric
down-conversion,'' \prl {\bf 76}, 4340 (1996).

\bibitem{11} A. G. Kofman and G. Kurizki, ``Quantum Zeno effect on
atomic excitation decay in resonators,''  \pra {\bf 54}, R3750
(1996).

\bibitem{12} A. G. Kofman, G. Kurizki, and B. Sherman, ``Spontaneous
and induced atomic decay in photonic band structures,''  \jmo
{\bf 41}, 353 (1994).
\bibitem{har98}
G. Harel, A. G. Kofman, A. Kozhekin, and G. Kurizki,
``Control of Non-Markovian Decay and Decoherence by Measurements and
Interference'',
Opt. Express {\bf 2}, 355 (1998).
\bibitem{mos72}
H. E. Moses, ``Exact electromagnetic matrix elements and exact
selection rules for hydrogenic atoms'', Lett. Nuovo Cim. {\bf 4}, 
51 (1972).

\end{references}
\end{document}